# 'Astronomy' or 'Astrology'

*a brief history of an apparent confusion*

A.Losev
Bulgarian Academy of Sciences, 11 G.Bonchev str., Sofia 1113, Bulgaria

**Abstract** The modern usage of the words astronomy and astrology is traced back to distinctions, largely ignored in recent scholarship. Three interpretations of celestial phenomena (in a geometrical, a substantialist and a prognostic form) coexisted during the Hellenistic period. From Plato to Isidore of Seville, the semiotic contrast is evidenced and its later developments are sketched. The concept of astronomy is found to be rather constant and distinct from changing views about astrology.

The contemporary cultural context allows to easily distinguish between astronomy and astrology. When needed, some discourse on physics is wedged between the two and it contrasts them, bringing support for the first but not for the second. This strategy turns out to be problematic in earlier times as an inverted situation appears there: physics founds astrology while astronomy is taken to be purely hypothetical. Language considerations point that today's astrology has appropriated the name of its founding knowledge. A statement that before the Modern Times no clear difference was made between astronomy and astrology is perhaps trivial but its explicitation is not really straightforward. Three conceptualisations of the celestial realm are found under the two names, which breeds complications and confusions.

In ancient texts sometimes one (or the other) word is used for both disciplines but no evidence appears for any inversion of the two names. This suggests that our word usage is not a convention but rather the outcome of an unstated tradition and the alleged indistinction might be only lexical. For the scholastics or even in the early Middle Ages the existence of two words implies existence of two realia and for all concerned the 'right' semantic coordination was not a problem. The person involved in celestial science is always an "astrologer" as if the *nomos* was among the stars themselves while their *logos* is a knowledge, needing an agent. Indeed the figure of the astronomer with his appellation is a late comer. Writing his monumental studies in history of science, Pierre Duhem chose to oppose two kinds of practitioners of celestial science and labelled them either 'astronomes' or 'physiciens' restricting the traditional 'astrologer' to superstitious astrology. Useful as it was, this tripartite division was just a methodological one, relying on contemporary views and word usage. In order to distinguish the physical from metaphysical content, or rather positivist phenomenism from metaphysical fancy he proposed 'saving the phenomena' as a slogan under which the 'astronomes' were seen to be laboring[1]. However the expression appears to be of a rather late coinage, just as the 'astronomer' and his whole reconstruction might seem to be somewhat arbitrarily imposed.

Classical studies since the 19th century have asserted indiscriminately equivalence between 'astronomy' and 'astrology'[2] even if lots of cases, read with regard to intention and content, just as Duhem did, disagree with this affirmation. The two words could be found to denote different disciplines and many ancient writers, at least those concerned by the distinction, used them knowingly. From Plato to Kepler, the coexistence for two millennia of a synonymic pair with similar word form would be a puzzling fact and just one occurrence of contamination seems to have been recorded.[3]. At the closing of the middle ages for rhetorical or ideological purposes the confusion might have been willful, betraying indeed a rather clear grasp of the issue. Not only historians but translators also have often made 'correct' substitutions, relying on their own judgment and thus obscuring further the distinction which is still present in the original texts.

---

[1] 'Σοζειν τα φαινόμενα' became the title of Duhem's book (1908) but the historical adequacy of the locution has been contested (Goldstein 1997).

[2] Asserted in various reference works such as Daremberg and Saglio's *Dictionnaire,* Lewis and Short *Latin Dictionary* or Smith et al., *A Dictionary of Greek and Roman Antiquities.* A Seminal paper by Pines (1964) challenged the received view. Recently Bowen (2007) considered an earlier instance, bypassing however the word usage.

The mastering of the calendar in times, when almost nobody could write and few people were able to count up to ten, has probably been deemed a prodigy. It embodies a precise foreknowledge of the sun's observable behaviour and, albeit imprecisely, relates it to sesonal happenings in nature. Different extrapolations were bound to appear. Babylonian scribes have left a remarkable collection linking the 'day to day' configuration of the nocturnal sky to various earthly happenings and it is just a small testimony for their obsession with any kinds of omens. Recordings of the form 'when $x$, occurrence of $y$' were accumulated, their content ranging from the trivial to the impossible. Exhaustion, as the degree zero of method, is not absurd in a world supposed to be finite - the spirit is truly positive even if it is also totally uncritical. For cyclic phenomena discovering their periods amounts to complete knowledge. Lack of causality, however, is a negative fact which practice does not reveal, so the Babylonians achieved actually a prediction of celestial omens but not of their apodoses, which remained as lists of precedents. In a similar fashion the Greeks composed their parapegma – meteorological recordings for each day of the year, but, understandably, as weather forecasters they did not achieve any success. Explaining the failure lead them to accept a difference in essence between the sublunar world and the higher realm. Thus Aristotle's decision to prescribe separate sciences for them eludes the problem by dividing it - a seemingly Cartesian gesture. However this splitting produced as a side effect what came to be known as astrology.

### *1 Plato or Aristotle: astronomy or astrology*

Plato's discussion of the disciplines in the *Republic (527d-530d)* includes the statement that geometry starts with planar figures and next it proceeds to solids and their movements which is properly the concern of astronomy (αστρονομία). The beginning of Aristotle's *Physics* neatly confronts Plato's conception: the enumeration *(194a7)* of 'sciences inverse of geometry' runs through optics, harmony and astrology (αστρολογία). There is no doubt about his meaning, as various translations unanimously testify, but using a different word emphasizes the difference. Plato goes on by mentioning how astronomy could be useful for navigation but points that we should be concerned with 'genuine astronomy'. Symmetrically, Aristotle *(Post. Anal. I.13, 79a)* remarks that astrology is both 'nautical and mathematical' and here as elsewhere he uses expressedly that word.'. A similar distinction is also found in Xenophon's *Memorabilia* when he incidentally remarks that travelling needs a certain 'practical knowledge of astrology' while knowing the movements outside the earthly orb is 'knowledge of astronomy' *(Mem. IV..7, 4-5)*. In Plato's works the word astronomy occurs at least twenty times, but his texts never had a role comparable to Aristotle's and it is only with the Neoplatonists, some time after Ptolemy, that his term regains a wider circulation. Porphyry wrote then an *Introduction to astronomy* and following his mentor's usage he mentioned that Pythagoras had learned 'geometry and astronomy' *(Vita Pyth*.11). This usage was severly eclipsed by Aristotle's teachings: Eudemos' *History of Astrology* had appeared in his lifetime and the term was adopted by all Peripatetics and the later Stoics.

The first explanation of the sun's movement as resulting from two rolling circles was apparently proposed within the Pythagorean school but Plato has been credited as the author of a full blown programme. According to Simplicius, he has proposed that the wandering of the planets is only apparent while their true movements are just a combination of uniform circular rotations. For this step from the phenomenal to the noumenal Plato adduced arguments and restrictions appealing to perfection, divinity and other ideological bias. Eudoxus' system came as a first realisation of the proposal, an event notable enough to provide a watershed between astronomy and astrology. Aristotle took to reintepret realistically and quasi-physically the construction that was generated theoretically - language itself reminding us here what was its origin[4]. Rather symptomatic, it was not done in the books 'about the heaven' or 'physics' but in the *Metaphysics*. Knowledge for Aristotle is a knowledge of causes and movement needs one. In order to build a mechanically causal explanatory model he introduced a few more 'unrolling' spheres which allowed to avoid unwanted transmission of movements. The centre of the system, which was originally just a geometrical point, has gained the status of a most important place in the universe. However around that time it became known that a combination of epicycle and deferent offered the best explanation which included rotations about different points. Awareness that is equivalent to eccentric orbits may have occurred to Hipparchus or somebody else and thus Aristotle's view clashed openly with the astronomic programme. A compromise was sought by declaring that models which are not strictly geocentric are just hypothetical or fictional. The better fit to observational data was devaluated and 'saving the phenomena' became the catch phrase for it. In this unfortunate category went Herakleides' semi-heliocentric model, Aristarchus' system and much later Copernicus' as presented in the Wittenberg interpretation. The physics invented by Aristotle took enough hold of reality to combat the earlier geometry and claim to be true. Actually it was only Kepler who conceived the *New Astronomy, Based upon Causes* as it was announced in the title of his book *(see below note 15)*. Indeed the causes are accounted for in Newton's mechanics which reproduces easily the phenomenology of the solar system. But even Newton refused to feign some hypothesis about the cause of gravity. The issue was solved later by introducing material fields, the same idea being already upheld by stoic thinkers who boldly asserted that 'causes are bodies'. Peripatetism and stoicism strongly favored substantial-causal explanations and geocentrism remained despite the clash with astronomical data.

---
[3] 'astrolomie' in Marco Polo's *Il Milione* (chap. cxxxix); the word being used for a man who made a prediction

The debate about celestial events extends to their consideration in the sublunar orb: even if it was heterogeneous the World is still a whole. Causal interaction, when viewed qualitatively, can be traced indefinitely far and this is what the fatalistic stoics did. The difference between the effects of the Sun and Moon and those of the other planets is only in degree, not in essence, and no reason to exclude them from consideration can be adduced. An other principle was upheld to cut the endless causal interactions – the self-evident freedom of the will. The occasion for this development was the coming into fashion of the Babylonian divinatory practice. The signs of gods' will, which they read, would be reified into astral influence by Greek thinking.

*2 The Babylonian connection*

---

[4] It would be almost a tautology to point that astronomy is the first 'theoria' – a way of seeing. Aristotle approach is meta-physic as he proposed to explain what is seen: a separate realm with its own laws. Nature, or 'physis', for him consists in generations and corruptions explained by the four elements but above the Moon there is a fifth substance. One is tempted to describe the appropriate science, astrology, which inevitably relies on earthly logic and analogies, as literally super-natural or at least para-physical (the situation is further complicated by viewing the soul also as a substance; interestingly, renaissance alchemy will be called sometimes 'astronomia inferior').

It is a mainly matter of speculation what Plato or Aristotle (and Eudoxus)[5] knew about Babylonian lore as its appropriation becomes perceptible only after Alexander's conquest. This is indeed the problem: why did this foreign practice come to proeminence so lately? Obviously, it is the conjuncture of accumulated knowledge and a flow of new information which provides a solution. This amounts to agree with a conclusion which, despite its numerous statements, still comes as surprise: astrology, as we know it, has been invented by the Greeks. Historical investigations lead to this view[6] but also consideration of its own working and valuations[7].

The first attested linking of an individual's birth with astral recordings – a horoscope- is in cuneiform writing, on a tablet dating from 405BE when Greeks were already speculating on astral matters. Keeping to the contemporary usage we could ask of what kind were the celestial concerns of Babylonian at that times. There is an obvious contrast but it would be totally unjust to assert that they were purely astrological. Rather the obverse, one may state that they actually discovered astronomy; it is commonly admitted that Pythagoreism developed into mathematical science and obviously a similar process could lead to the appearance of astronomy on the baylonian kingdom[8]. The indebtedness to superstition, religion or myth would not be greater that the one inherited from Plato and Aristotle who took for granted the divinity of planets. Tabulating astronomical data, using ecliptic coordinates, a numerical system based on 60 with a marking for zero are elements of a discipline which surpasses in rigor and precision most Greek endeavours. Stellar data are the main content of Babylonian horoscopes while their interpretation is sketchy, relying on annals and tradition (*Rochberg F., 1998)*. A transfer to Greece would mean to carry over this part which is algorithmically irreducible. The general idea however is easily transmissible and the Greek implemented it with their own means. A similar instance would be the development leading from common law to roman law, both being practices to achieve a particular aim. Of course this inversion - astronomy being Babylonian while astrology is Greek - does not really matter, except for the perspective which the participants could have had. For historical purposes one might just as well agree with Philo of Alexandria who says that Chaldeans invented both 'astronomy and genethlialogy'[9].

So, celestial science in Ptolemy's time is conceived in three main interpretations: the first, a Pythagorean one, as geometry, the second one, physicalist and substantial was inspired by peripatetism and a last one, prognostical, attributed to the Chaldeans. The earlier 'astronomy' steps back before the later 'astrology', while the newcomer still needs a name. It is called descriptively by referring to its alleged originators, the Chaldeans, or known as 'apotelesmatics' and, more particularly, as 'genethlialogy' or 'katarkhe'. For Latin authors, and for anybody not involved in this, the distinction between Pythagorean, peripatetic or Babylonian views would have been rather elusive. What is more, Babylonian tables allowed preparation of horoscopes and celestial prognostication without any grasp of astronomy. Any 'astronomer' could and did the same, so the profession denomination is 'astrology' and correspondingly its practitioner – the astrologer. Before the first century latin did not use 'astronomy' (exception being made perhaps for the *Astronomica* of the mysterious Manilius), as the majority of Greek had adopted astrology. The former term was still currently used as witnessed by the texts of Theon of Smyrna or the data collected by Diogenes Laertius[10]. Sextus Empiricus in writing against the doctores of his day notes that 'Chaldeans call themselves mathematici or astrologi' and attacks their astrology or 'mathematical art differing from arithmetic and geometry' and different from 'the prognostics of Eudoxus and Hipparchus, which some call astronomy' (*Adv. math.* V 1-2).

---

[5] The *Timaeus* offers privides grounds for some acquaintance to be acknowledged while Aristotle's remarks remain in a naturalistic vein; according to an uncorroborated remark in Cicero, (*De Div.*, ii, 42, 87), Eudoxus demanded that 'no credence should be given to the Chaldeans, who predict and mark out the life of every man according to the day of his nativity'.

[6] Neugebauer O., *The Exact Sciences in Antiquity*. NY Dover publications, 1969. p.80; Pingree D., *Astrology* in The Dictionary of the History of Ideas; Rochberg-Halton F., Elements of the Babylonian Contribution to Hellenistic Astrology, *J of the American Oriental Society*, 108 (1988), No. 1, p.51

[7] Beck R., *A brief history of Astrology*, London, Blackwell, 2006

[8] Today Babylonian mathematics is understood to be mostly arithmetic but, rather curiously, Josephus wrote in his mythical account of the *Jewish Antiquities* that Chaldeans learned from Abraham 'arithmetic and astronomy' (I.8.2 (166)), the usual pair of 'geometry and astronomy' appearing elsewhere (I.3.9 (106)).

[9] *De peregrinatione Abrahami* 33.178; in 35.194 genethlialogy is explained briefly

### *3 Ptolemy's shuffle*

The larger flow of history has given a Ptolemy a distinct place as he has remained for a millennium the authority in astronomy and even longer in astrology. His achievement appears to be not so a novelty as a reconfiguration. Instead of the dilemma describing/explaining his work brings to the front knowledge in the form of prediction – it can be only more or less exact. Describing the celestial movements is apodictic while tracing their causes or effects is just probabilistic.

The eclecticism of the zeitgeist is perceptible in Ptolemy's writings which comprise both platonic astronomy and peripatetic-stoic physics. Aristotelian astrology has always been something like an astral twin of sublunary valid knowledge and obviously no room is left for it in this mix. The return to a Pythagorean tradition is obviously and the avoidance of the Aristotelian term marked[11]. The dual hierarchy of Aristotle cosmos is replaced by a fourfold schema built on oppositions from the categories 'immaterial' and 'invisible' Theology is the science about the immaterial and invisible, mathematics is about the immaterial and visible while physics is about the material and visible. The material and invisible, which corresponds to the soul, is subsumed in physics and this imbalance reveals that the really meaningful distinction is between ideal and material.

The four books or *Tetrabiblos,* devoted to what is today's astrology, were known as Ptolemy's *Apotelesmatics* which is his own preferred term, explained as the prognostication by astronomy. In the celestial realm predictions are strictly true while anywhere else they are only probable - for meteors or individual predictions. But a continuously distributed probability erases the opposition between sublunar and higher realms and thus invalidates the Aristotelian difference between astrology and physics. Lacking a proper content 'astrology' can be used for the founding and explaining of the astral influences as previously done by physics. And this is what really happens, but much later, when Aristotelian science is fully discredited. For the moment 'astronomical prediction' or some such paraphrasing is commonly used as it is mostly taken in the same restricted sense as 'astrologer'. Some interest about the star patterns when they are devoid of divinity and without a look for their effects would have been odd indeed. So a first modern looking definition of astrology as judging or predicting by the stars appears to have been given by more pragmatic Arabian commentators *(Pines, 347)*

### *4 Fast forward*

Since late antiquity the quadrivium provided a context which unambiguously identifies astronomy independently of the word used. Mathematics, already in a restricted sense, included two proper subdisciplines, arithmetic and geometry and they had as counterparts music and a celestial science. Varro and Martianus Capella still called it 'astrology' but

---

[10] 'Astronomy' occurs at least in 4 instances, with more than 10 for 'astrolog-er/y'; Diogenes Laertius collated various sources and so the word usage can appear as inverted.

[11] In the *Almagest* none of the words appears; in the *Tetrabiblos* 'astronomy' is used 6 times and, as Feke notes, its only other appearance is in the *Harmonics* where it is defined as a mathematical science, cf. Feke J.(2009), *Ptolemy in Context*, p.153.

Cassiodorus adheres strictly to 'astronomy' even if he might be referring to Varro's *De Astrologia*. The existence of two words assured medieval authors that there are two things and they were able to provide an educated guess as apparently Alcuin or Hugo of Saint Victor[12] did and they discussed separately the geometrical Pythagorean science from its more substantial variants. In early 9th century Martin of Laon enumerates the disciplines from the quadrivium ending by "astronomy to which cling astrology and medicine"[13]. The same disposition is found much later when university education has been instituted. Aristotle is taught at the theological faculty while the phenomena-saving astronomy and its astrological and medicinal continuation had its place at the faculty of medicine. Galileo still had to teach them there. Aristotle's texts mentioning astrology became known to Western scholars a few centuries after they had learned from the Arab tradition about 'judging by stars'. Liber de *Astronomice judicandi* by Roger of Hereford is an early example *(ca.*1184) of astrological treatise presented with the words that Ptolemy might have used. Improving the calendar has been a prime interest of ecclesiastics and there has been an awareness that astronomical tables either Arabian or Babylonian offer valid data only for the location where they are computed, so any prognostication needs astronomy as its precondition.

Isidore of Sevilla had compiled his *Etymologies,* including a comment about *The difference between astronomy and astrology* which surely would not have been there if it was not in some earlier text. It is worth noting that his definition of astronomy reproduces a wording by Cicero who was writing about astrology, so Isidore, or somebody before him, knew enough to transpose this usage[14]. Remarkably, Isidore goes on and makes a further distinction which divides its topic in three parts. After separating astronomy from astrology, he adds that the later is 'partly natural, partly superstitious', which would correspond to an Aristotelian and to a Babylonian concept. The religious qualification here etymologically speaks about 'standing- over' or 'supernatural', which is indeed what Chaldean science was. A "natural astrology" would have been for a peripatetic something of a contradictio in adjecto, just as "celestial physics"*,* used in a book's title[15] much later by Kepler. Nevertheless the same text reappears elsewhere[16], and the *Etymologies* remained influent through the Middle Ages, transmitting an understanding achieved already at the start of Hellenism.

It seems safe to conclude that through the ages people who used the word astronomy knew what they were talking about. Late medieval and Renaissance writers sometimes stretched the term to cover most of what is astrology but such a rhetorical strategy would not have been possible without a prior knowledge of the difference[17]. Acknowledging the history hidden behind the name astrology leads to a clearer grasp of the ambiguities in its usage.

---

[12] Alcuin *PL101:947* Astronomia lex astrorum, qua oriuntur et occidunt astra. Astrologia est astrorum ratio et natura et potestas, coelique conversio Hugo St Victor *PL176:756* astronomia de lege astrorum nomen sumpsit, astrologia autem dicta est quasi sermo de astris disserens. Nomos enim lex et logos sermo interpretatur. (cf *Pines*).

[13] Martin of Laon /Hiberniensis (819-875) Physica in quattuor divisines partitur id est arithmeticam, musicam, geometriam, astronomiam quibus adherent astrologia, medicina. in *Insular Latin Studies* ed. M.W. Herren, Toronto, 1981

[14] *Etym 3.27*: Astronomia caeli conversionem, ortus, obitus motusque siderum continet, in the enumeration of disciplines by Cicero:.. 'Astrologia, caeli conversio, ortus, obitus motusque siderum' *(De Oratore ii.42)*.

[15] *Astronomia nova, Αἰτίολογητοσ, seu Physica coelestis* translated as "New Astronomy, Based on Causes, or Celestial Physics" (1609)

[16] PL 90, 908D, in *Dubia et Spuria* of Bede astronomy and astrology are named as two of the six parts pertaining to physics and then the same text is reproduced

[17] Since the end of the 13th century there has been a discussion how much of astrology is 'licit': the Church and the secular power maintained conflicting opinions which were further complicated by the humanists' views during the Renaissance. As a defender of astrology Pierre d'Ailly went as far as to write about 'astronomy falsely known as astrology' in his *Tractatus de concordantia theologie et astronomie* while Pico della Mirandola's *Disputationes adversus astrologiam divinatricem* dealt it a nearly fatal blow.

Pleonastically looking qualifications as 'divinatory' or 'judicial astrology' are witnesses of the distinction from a 'physical' or 'natural astrology', an early attempted science which became sidetracked.

**References**

< old texts, except where noted, are to be found at http://www.archive.org /details/texts>

Aristotle. 1871. *Aristotelis opera*, ed. I. Bekker, Berlin:Academia Regia Borussica
Beck, Roger. 2006. *A brief history of Astrology*, London: Blackwell
Bowen, Alan. 2007. "The Demarcation of Physical Theory and Astronomy by Geminus and Ptolemy", *Perspectives on Science* 15: 327
Cicero. *Opera* at The Latin Library <http://www.thelatinlibrary.com/cic.html>
Daremberg C., Saglio E. 1919. *Dictionnaire des Antiquités Grecques et Romaines,* Paris:Hachette <http://dagr.univ-tlse2.fr/sdx/dagr/feuilleter.xsp ?tome=1&partie=1 &numPage=488>
Diogenes Laertius. 1972. *Lives of Eminent Philosophers*. Translated by R.D. Hicks (1925). Cambridge: Harvard University Press.
Duhem Pierre 1908 Σοζειν τα φαινόμενα *Sozein ta phainomena. Essai sur la notion de théorie physique de Platon à Galilée*, Paris: Hermann
Josephus Flavius, *The Judean Antiquities,* <http://pace.mcmaster.ca/york/york/showText? text=anti>
Feke, Jacqueline. 2009. "Ptolemy in Context", PhD dissertation, Toronto
Goldstein B., 1997, Saving the phenomena: the background to Ptolemy's planetary theory, *J.His.t Astronomy* 28 (1997) p1
Lewis C., Short C. 1879. *A Latin Dictionary*. Oxford: Clarendon Press.
Neugebauer, Otto. 1969. *The Exact Sciences in Antiquity*. NewYork: Dover Publications.
Pingree, David. 1968. Astrology in *The Dictionary of the History of Ideas*. NewYork: Scribners
Patrologia Latina (PL). 1844**.** *Patrologiae cursus completus:* vol.83 (Isidorus Hispanesis), vol.90 (Beda Venerabilis), vol.101 (Alcuinus), (Hugo de S. Victore), vol.176, ed. J.-P. Migne, Paris
Philo of Alexandria. 1830. *Philonis Judaei opera omnia*, ed. M. Richter, Leipzig: Schwickerti
Pierre d'Ailly. 1414. *Tractatus de concordantia theologie et astronomie* <http://warburg.sas.ac.uk/pdf/fah1620p.pdf >
Pines, Shlomo. 1964. "The Semantic Distinction between the Terms Astronomy and Astrology according to al-Biruni" *Isis*, 55: 343-349
Polo, Marco. 1928. *Il Milione,* ed L. Benedetto, Firenze (L. S. Olschki)
Plato. 1903. *Platonis Opera*, ed. J. Burnet. Oxford: Oxford University Press.
Rochberg, Francesca. 1998. *Babylonian horoscopes*, Trans. Am. Philos. Soc. 99 (1998), 1.
Rochberg-Halton, Francesca. 1988. "Elements of the Babylonian Contribution to Hellenistic Astrology", *J. of the American Oriental Society*, 108 (1988), No.1, p.51
Sextus Empiricus. *Sexti Empirici Viri Longe Doctissimi Adversus Mathematicos,* G. Hervetus, Paris, 1569
Smith W., Wayte W. and Marindin G. 1890. *A Dictionary of Greek and Roman Antiquities*, London: John Murray.
Xenophon. 1921. *Xenophontis opera omnia*, Oxford: Clarendon Press.